\documentclass[%
 reprint,
 superscriptaddress,
 amsmath,amssymb,
 aps,
prb,
]{revtex4-2}

\usepackage{color}

\usepackage{graphicx}
\usepackage{dcolumn}
\usepackage{bm}

\usepackage{color}

\newcommand{\ket}[1]{\left|#1\right\rangle}

\begin{document}

\preprint{APS/123-QED}

\title{Error per single-qubit gate below $10^{-4}$ in a superconducting qubit}
\author{Zhiyuan Li}
\thanks{Z. Li and P. Liu contributed equally to this work.}
\affiliation{Beijing Academy of Quantum Information Sciences, Beijing 100193, China}
\author{Pei Liu}
\thanks{Z. Li and P. Liu contributed equally to this work.}
\affiliation{State Key Laboratory of Low Dimensional Quantum Physics,
Department of Physics, Tsinghua University, Beijing 100084, China}

\author{Peng Zhao}
\affiliation{Beijing Academy of Quantum Information Sciences, Beijing 100193, China}

\author{Zhenyu Mi}
\affiliation{Beijing Academy of Quantum Information Sciences, Beijing 100193, China}

\author{Huikai Xu}
\affiliation{Beijing Academy of Quantum Information Sciences, Beijing 100193, China}

\author{Xuehui Liang}
\affiliation{Beijing Academy of Quantum Information Sciences, Beijing 100193, China}

\author{Tang Su}
\affiliation{Beijing Academy of Quantum Information Sciences, Beijing 100193, China}

\author{Weijie Sun}
\affiliation{Beijing Academy of Quantum Information Sciences, Beijing 100193, China}

\author{Guangming Xue}
\affiliation{Beijing Academy of Quantum Information Sciences, Beijing 100193, China}

\author{Jing-Ning Zhang}
\email{Corresponding author: zhangjn@baqis.ac.cn}
\affiliation{Beijing Academy of Quantum Information Sciences, Beijing 100193, China}

\author{Weiyang Liu}
\email{Corresponding author: liuwy@baqis.ac.cn}
\affiliation{Beijing Academy of Quantum Information Sciences, Beijing 100193, China}

\author{Yirong Jin}
\affiliation{Beijing Academy of Quantum Information Sciences, Beijing 100193, China}
\author{Haifeng Yu}
\affiliation{Beijing Academy of Quantum Information Sciences, Beijing 100193, China}


\date{\today}

\begin{abstract}
Implementing arbitrary single-qubit gates with near perfect fidelity is among the most fundamental requirements in gate-based quantum information processing. In this work, we fabric a transmon qubit with long coherence times and demonstrate single-qubit gates with the average gate error below $10^{-4}$, i.e. $(7.42\pm0.04)\times10^{-5}$ by randomized benchmarking (RB). To understand the error sources, we experimentally obtain an error budget, consisting of the decoherence errors lower bounded by $(4.62\pm0.04)\times10^{-5}$ and the leakage rate per gate of $(1.16\pm0.04)\times10^{-5}$. Moreover, we reconstruct the process matrices for the single-qubit gates by the gate set tomography (GST), with which we simulate RB sequences and obtain single-qubit fedlities consistent with experimental results. We also observe non-Markovian behavior in the experiment of long-sequence GST, which may provide guidance for further calibration. The demonstration extends the upper limit that the average fidelity of single-qubit gates can reach in a transmon-qubit system, and thus can be an essential step towards practical and reliable quantum computation in the near future.
\end{abstract}

\maketitle

\section{\label{sec:level1}Introduction}

Reliability is an unavoidable crux in the quest for beyond-classical computational capabilities. As to the circuit-based quantum computation, improving the reliability of entire computational tasks is decomposed into a series of subtasks, among which implementing high-fidelity single-qubit gates is an important component. For example, both the noisy intermediate-scale quantum (NISQ) application and the fault-tolerant quantum computation, i.e. the near-term and the ultimate goals of the circuit-based quantum computation, make requirements on gate fidelities that exceed the state-of-the-art values. As a result, considerable efforts have been invested in realizing high-fidelity single qubit gates in leading platforms for quantum information processing.
As to the superconducting quantum computation, great progress has been made over the past two decades, including the realization of accurate and precise quantum gates. The single- and two-qubit gate errors in transmon qubit are below $10^{-3}$ and $10^{-2}$ \cite{jurcevic2020,somoroff2021,sungRealizationHighFidelityCZ2021a,stehlik2021PRL,kandala2021PRL,wei2022PRL,acharyaQEC2022,bao2022PRL} respectively, and the single-qubit gate error in fluxonium qubit is below $10^{-4}$ owing to a millisecond coherence time \cite{somoroffMillisecondCoherenceSuperconducting2021}. To further improve the single-qubit gate fidelity, identifying the nature of the
dominant errors is particularly
important for improving performance.and the fidelity of single-qubit can be seen as the upper boundary of two-qubit gate.

One of the essential requirements for the reliable implementation of circuit-based quantum computation is a sufficiently large ratio between the coherence time and the gate length. To increase this ratio, one prevalent way is to increase the coherence times of the quantum devices, including the energy relaxation time and the dephasing time. As to the superconducting transmon system, the detrimental impact of the two-level-system (TLS) defects in dielectrics on the energy-relaxation time has been extensively studied~\cite{melville2020APL, murray2021MSaERR, woods2019PRA}. To relieve 
this impact, two alternative directions have been explored, one is to suppress the coupling between the TLS defects and the transmon by optimizing the geometry~\cite{martinis2022nQI}, and the other is to lower the density of the TLS defects by appropriate materials and recipes~\cite{Place2017nc,wang2022nQI}. 
On the other hand, environmental noises, such as the fluctuation of magnetic flux and the residue of the thermal photons, can cause qubit dephasing and thus decrease the dephasing time \cite{bialczak2007PRL,goetz2017PRL, tomonaga2021PRB, yan2018PRL}. The solution might be to keep the transmon-qubit well separated from its environment, e.g. mitigating its couplings to the drive lines, the readout cavity or the neighboring qubits. However, this is obviously in contradiction to the implementation of fast qubit control, which requires the qubit apt to be driven. As a result, balancing the coherence times and the gate length becomes particularly important for the implementation of high-fidelity single-qubit gates.
Besides, to mitigate the impact of the spurious reflection signal due to impedance mismatch in the line, a short buffer should be added after each single-qubit operations to avoid residual pulse overlap \cite{chow2009PRL}. 

In this work, we design and fabricate a superconducting device consisting of superconducting transmons, which are of long coherence times and apt to strong drivings. With this device, we implement a set of single-qubit gates, constructing from the $X_{\frac{\pi}{2}}$ gate and the virtual Z gates. The average gate fidelity, as well as the fidelities for $\pi/2$-pulses, is benchmarked to be higher than $99.99\%$, exceeding the state-of-the-art record in superconducting transmon-qubit systems. We also analyse the sources of the residual errors, including the incoherent error the leakadge rate per Clifford gate. As a cross validation, we experimentally obtain process matrices for the identity and $\pi/2$-pulses with the gate-set tomography. Besides extending the computational upper limit of a transmon-qubit processor, our experiment also indicates that the bottleneck to further increase the reliability might be to suppress the non-Markovian effect.

\section{\label{sec:level2}Experiment}

\begin{figure}[t]
	\includegraphics[width=0.48\textwidth]{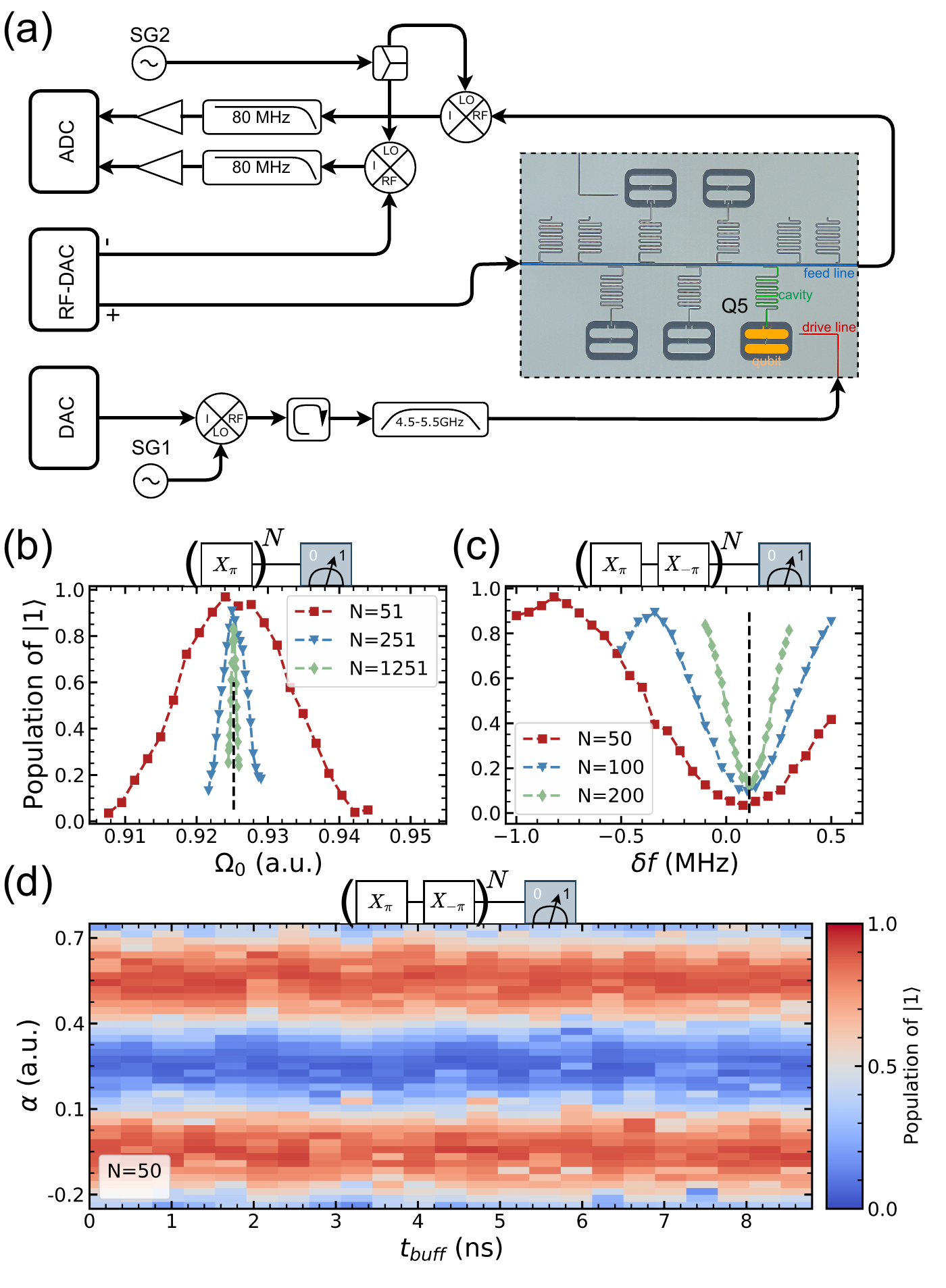}
	\caption{\label{fig:gate_characterization}(a) The optical image of the qubit device and the schematic of the experiment setup. 
    (b) Calibration of the pulse amplitude $\Omega_{0}$ for the $X_{\pi/2}$ gate. 
    The inset shows the pseudo-identity sequence. (c) Optimize the detuning of the pulse. The pseudo-identity sequence is shown in the inset. The curves show the excited state population as a function of detuning $\delta{f}$ for N = 50, 100, 200 pairs. (d) $X_{\pi}$ and $X_{-\pi}$ gate sequence to check the trailing edge of the pulse. There is no population change along the duration of the buffer $t_{\rm{buff}}$ of the pseudo-identity sequence.
	}
\end{figure}

We implement high-fidelity single-qubit gates on a fixed-frequency transmon qubit fabricated with tantalum films~\cite{Place2017nc,wang2022nQI}. As shown in the insert of Fig.~\ref{fig:gate_characterization} (a), the qubit, labelled by $Q_5$, is embedded on a superconducting device consisting of five separate transmon qubits, each of which is coupled to a readout cavity sharing one transmission line. The $Q_5$ qubit is coupled to a microwave control line, which facilitates fast single-qubit operations. The transition frequency between $|0\rangle$ and $|1\rangle$ and the anharmonicity are $\omega_{01} = 2\pi\times4.631\:\mathrm{GHz}$ and $\Delta = -2\pi\times240 \:\mathrm{MHz}$, respectively. As to the coherence times, the energy relaxation time and the dephasing time are measured to be $T_{1} = 231$ $\mu$s and $T^E_{2}=204$ $\mu$s. These long coherence times indicate that the qubit is quite isolated from its external environment. Together with the electronics for the control and measurement system shown in Fig.~\ref{fig:gate_characterization} (a). At this coherence time level, single-qubit gates with the gate length of $20$ ns and single-shot measurement with high fidelities (see Supplementary Materials (SM)~\cite{supplementary}) can be achieved.
Specifically, we use the single-sideband (SSB) technology to modulate the control pulses, where the microwave signals are generated by a Digital-to-Analog Converter (DAC), a Signal Generator (SG1) and a mixer. To suppress local leakage and unwanted spurious signals, like the image or reflection of the control signals due to impedance mismatch, a bandpass filter and an isolator are introduced before the control signals access the qubit. The readout pulses are generated by a radio frequency-DAC (RF-DAC) with the sampling rate of 25 GS/s and fed into the fridge from the positive ($+$) port, while an out-of-phase signal from the negative ($-$) port is used as reference. Finally, the down-converted signals, going out of the fridge, are collected by an Analog-to-Digital Converter (ADC). 

In this experiment, we construct arbitrary single-qubit rotations with $X_{\frac{\pi}{2}}$ pulses and virtual Z gates~\cite{mckay2017PRA}. As the virtual Z gates can be treated as faultless, $X_{\frac{\pi}{2}}$ is the only gate that needs precise calibration. Here we implement the $X_{\frac{\pi}{2}}$ gate by a microwave pulse with a cosine-shaped envelope $\Omega(t) = \Omega_0\left(1-\cos(2\pi t/t_g)\right)$ with the gate length $t_g=20~{\rm ns}$. To suppress leakage and phase errors, we introduce the derivative reduction by adiabatic gate (DRAG) scheme~\cite{gambetta2011PRA, motzoi2009PRL, chen2016PRL} for the pulse envelope, i.e. $\Omega_{\rm DRAG}(t)=e^{i2\pi\delta f t}\left(\Omega(t)-i\alpha\frac{\dot\Omega(t)}{\Delta}\right)$ with $\Delta$ being the anharmonicity of the transmon qubit, where $\alpha$ and $\delta f$ are the DRAG weighting and detuning, respectively. Considering the fact that long periodic sequences can boost the sensitivity of certain coherent errors, we design and implement long-sequence calibration schemes to determine the exact values of the pulse amplitude $\Omega_0$ and the DRAG detuning $\delta f$. To be specific, $\Omega_0$ is calibrated with $N$ times of $X_\pi$ pulses, each of which is composed of two $X_{\frac{\pi}{2}}$ pulses. With $N$ being a large and odd number, we sweep the value of $\Omega_0$, with the optimum marked by a peak in the population of the $\ket{1}$ state, denoted as $P_1$. The calibration results are shown in Fig.~\ref{fig:gate_characterization}(b). The periodic calibration sequences for $\delta f$ consist of $N$ psudo-identity operators, each of which is composed of a composite $X_\pi$ pulse and its inverse gate, i.e. the $X_{-\pi}$ gate. As shown in Fig.~\ref{fig:gate_characterization}(c), the minimum in $P_1$ when sweeping $\delta f$ gives the opitimum of $\delta f$. The experimental results clearly demonstrate the increasing sensitivity as the sequence length, which is $\propto N$, increases. The final step is to determine the length of the buffer by what time the spurious signals can be neglected. In order to shorten the buffer time, we measure the trailing edge of the driving pulse by directly using the qubit~\cite{gustavsson2013PRL} and then correct the signal by pre-distortion. The efficacy of this technique is reflected in the experimental data shown in Fig.~\ref{fig:gate_characterization}(d). Starting from the $\ket{0}$ state, we implement a sequence consisting of 50 pairs of $\left(X_{\pi}, X_{-\pi}\right)$ pulses and measure the population in the $\ket{1}$ state. We collect experimental data by sweeping the duration of the buffer and the DRAG weighting parameter $\alpha$, with the variation of the latter providing an oscillating pattern . Although we observe no visible movement of the pattern as the buffer time increases, we still chose $t_{\rm buff}=2$ ns to guarantee there is no overlap between pulses, i.e. the gate length of a single $X_{\frac{\pi}{2}}$ gate is 22 ns.

\begin{figure}[t]
	\includegraphics[width=0.49\textwidth]{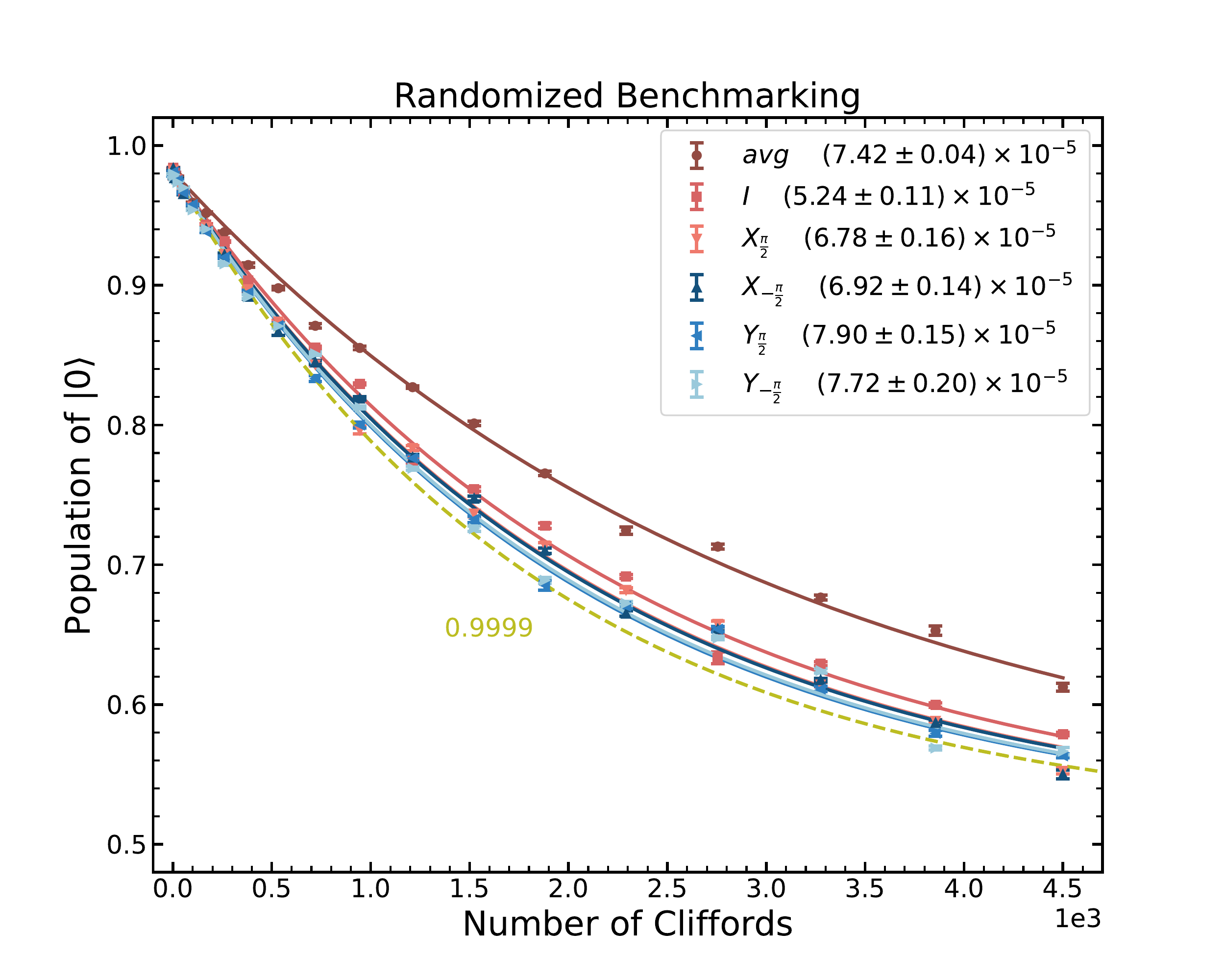}
	\caption{\label{fig:rb_exp} Average sequence fidelity as a function of the number of Clifford gates for the reference and interleaved RB. Each data point is averaged over 20 different random sequences, with the error bars being the standard error of the mean.    The average error rate, obtained from the reference RB, and the error rates of the identity and the $\pm\pi/2$ pulses are listed in the legend, which are all smaller than $10^{-4}$, and the uncertainties are obtained by bootstrapping. For comparison, the yellow dashed line indicates the exponential decay for the average gate fidelity of 0.9999.  
}
\end{figure}

The Clifford-based RB~\cite{emerson2007symmetrized, knill2008randomized}, as well as the interleaved RB~\cite{barends2014N}, is adopted widely to characterize the performance of quantum devices, which is approximately independent of errors in the state preparation and measurement (SPAM). Essentially, the RB experiment uses the fidelities of random pseudo-identity sequences, consisting of $m$ randomly sampled Clifford gates and their inverse, to estimate the average gate fidelity. With the original RB sequences as reference, the interleaved RB sequences have the the target gate, which should also be a Clifford gate, interleaved between the Clifford gates in the reference sequences, with the last inverse gate recalculated to guarantee pseudo-identity.  The waveforms corresponding to all sequences are collected first, and then the DAC output them sequentially. To get better statistical results, we generate 20 different random sequences for each value of $m$ and repeat each sequence 1024 times. The maximum number of Clifford gates is 4500 to achieve the credible precision ($\mathcal{O}(1/N)$). To partly resist the impact of the unstable readout and temporal drift of the system parameters, we perform the reference and interleaved RB sequences alternately with respect to the Clifford gate numbers $m$.


The measured average sequence fidelities for the reference and interleaved RB experiments are shown in Fig.~\ref{fig:rb_exp}, which are fitted by the exponential model $F_{\rm seq}(m) = Ap^m + B$, with $m$ being the sequence length, to obtain the reference and interleaved decay parameters, denoted as $p_{\rm ref}$ and $p_{\rm int}$, respectively. Here $A$ and $B$ are fitting parameters to accommodate the errors in SPAM and the last gate. The error per Clifford (EPC) is then extracted by $r_{\rm{clif}} = (1-p_{\rm{ref}})(1-1/d)$, with $d=2$ for a single qubit. Combined with the fact that in this experiment each Clifford gate consists of 2.2083 $\frac{\pi}{2}$-pulses on average, the average EPG is given by $r_{\rm avg}=r_{\rm clif}/2.2083=(7.42\pm0.04)\times10^{-5}$. To assess the performance of each of the single-qubit gates in ${\mathcal G}\equiv\left\{I, X_{\pm\frac{\pi}{2}}, Y_{\pm\frac{\pi}{2}}\right\}$, we continue to perform interleaved RB experiments, where the gate sequences are derived from those of the reference RB experiment by adding the specific gate right after each of the $m$ random Clifford gates. The EPG for the interleaved gate $G\in{\mathcal G}$ is given by $r_G=(1 - p_{\rm{\rm{int}}}/p_{\rm{ref}})(1 - 1/d)$. As shown in Fig.~\ref{fig:rb_exp}, the average EPG and those of specific gates in ${\mathcal G}$ are all lower than $10^{-4}$, with the EPG of the identity operator $r_I$ being the lowest and indicating the coherence limit of the transmon qubit.

\begin{figure}[t]
	\includegraphics[width=0.48\textwidth]{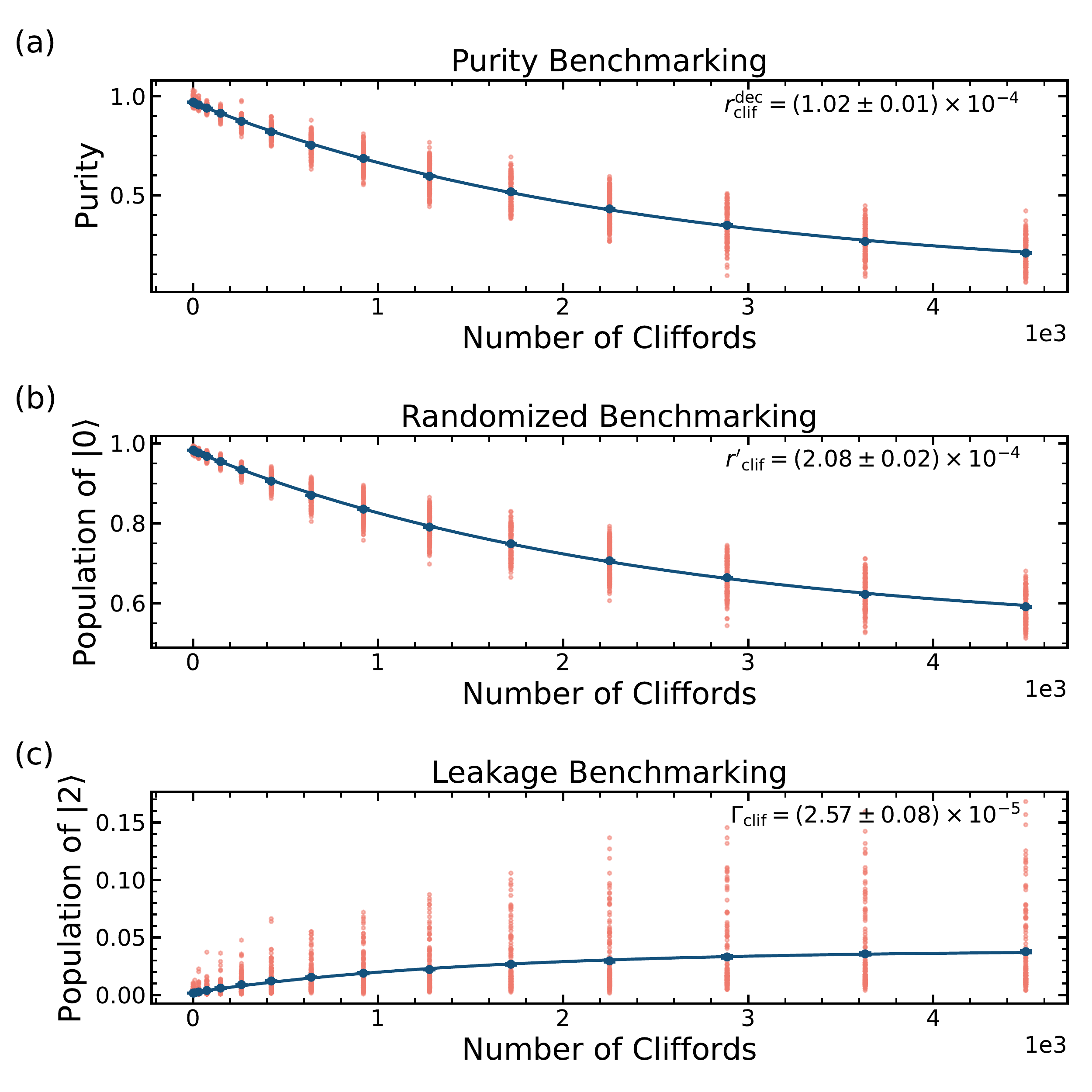}
	\caption{\label{fig:pb_exp}(a) The sequence purity averaged over 100 random sequences as a function of the number of Cliffords $m$ for the reference. (b) The sequence RB (100 averages) is a function of the number of Cliffords $m$ for the reference. (c) $|2\rangle$ state population versus the gate numbers showing accumulation of leakage with gate numbers. 
	}
\end{figure}

\begin{figure*}[t]
	\includegraphics[width=1.0\textwidth]{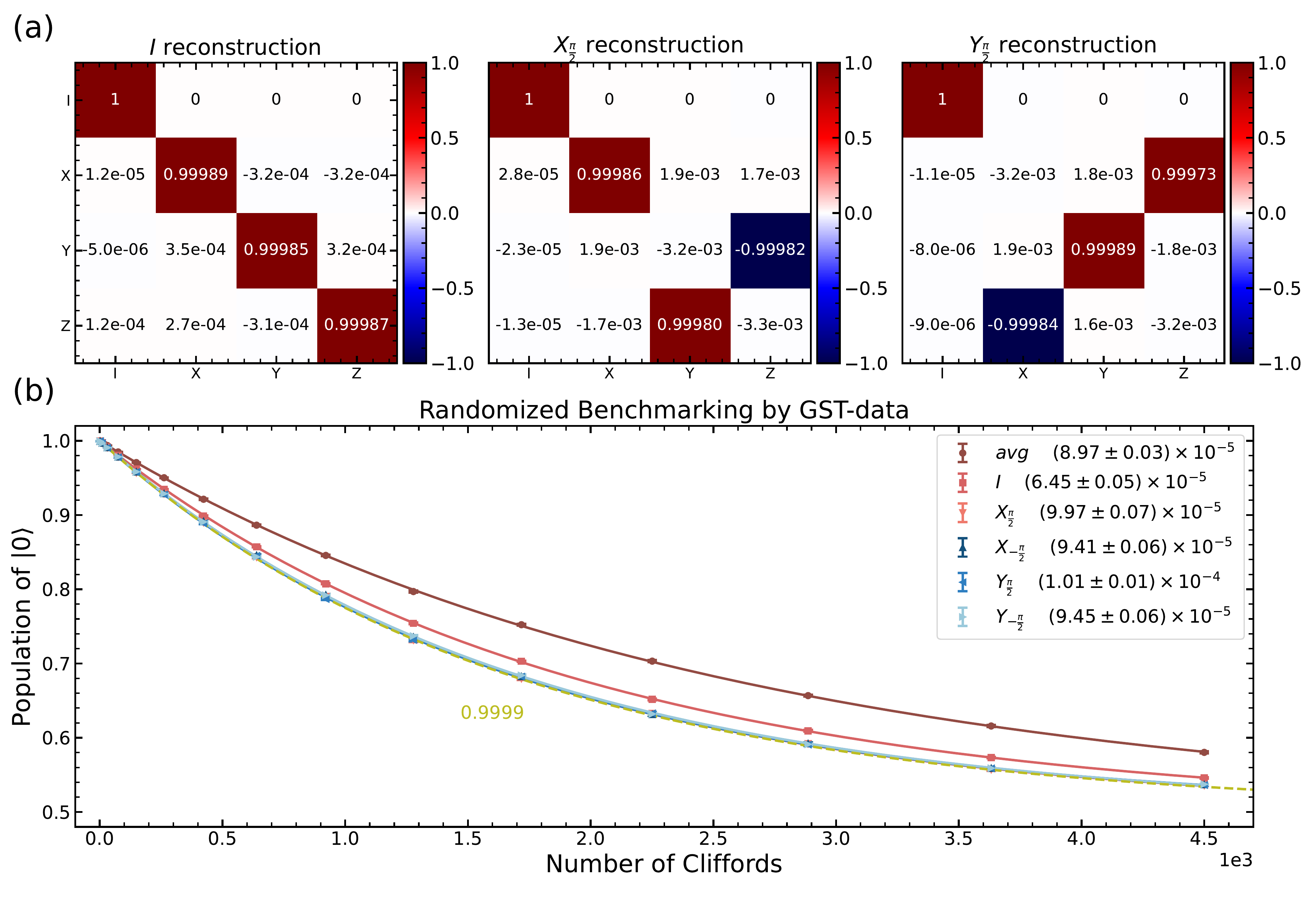}
	\caption{\label{fig:gst_mv_mat}
    (a) Reconstructing process matrix estimated by GST. The process matrix estimates of the $I$, $X_{\frac{\pi}{2}}$ and $Y_{\frac{\pi}{2}}$ gates are shown as superoperators on the basis of Pauli matrices, respectively.(b) RB data simulated using the gate set tomography $\mathcal{G}_0$ derived from experimental GST results. The reconstructing process matrix generates the single qubit Clifford group. Simulate the RB experiment and analysis the gate fidelity.
	}
\end{figure*}

To investigate error sources, we implement the purity benchmarking (PB) experiment~\cite{wallman2015NJP} and monitor the $\ket{2}$ state population, to characterize the imperfections caused by decoherence and leakage. The PB is an RB-based technique, which is delibrately designed to be insensitive to coherent control errors. The random Clifford gate sequences used in PB are generated in the same way as those in RB, with the final projective measurement replaced by quantum state tomography. The experimental data of the average sequence purity, i.e. $P_{\rm seq}(m)=\left\langle{\rm Tr}\left[\hat\rho^2\right]\right\rangle_m$ with the average taken over random sequences of length $m$, are then fitted by the exponential model $P_{\rm seq}(m)=Au^{m-1}+B$ to get the average purity decay parameter $u$, which gives a lower bound for the average decoherence EPC $r_{\rm clif}^{\rm dec}=(1-\sqrt{u})(1-1/d)$. To obtain the leakage rate per Clifford $\Gamma_{\rm clif}$,  we fit the measurement results of the $\ket{2}$ state population by a discrete rate equation $P_{\ket{2},m}=P_{\ket{2},\infty}\left(1-\exp\left(-\Gamma_{\rm clif} m\right)\right)+P_{\ket{2},0}\exp\left(-\Gamma_{\rm clif} m\right)$. With the experiment results in Figs.~\ref{fig:pb_exp} (a) and (c), the lower bound of the average decoherence EPC and the leakage rate per Clifford are estimated to be $r_{\rm clif}^{\rm dec}=\left(1.02\pm0.01\right)\times10^{-4}$ and $\Gamma_{\rm clif}=\left(2.57\pm0.08\right)\times10^{-5}$, respectively. The lower bound of decoherence EPG and the leakage rate per gate is $r_{\rm avg}^{\rm dec} = r_{\rm clif}^{\rm dec}/2.2083 = \left(4.62\pm0.04\right)\times10^{-5}$ and $\Gamma_{\rm avg}=\Gamma_{\rm clif}/2.2083=(1.16\pm0.04)\times10^{-5}$, with the former consistent with the expected EPG in the decoherence limit, i.e. $4.77\times10^{-5}$ given by numerically solving the master equation. Together with the averaged EPG obtained using the same set of measurement results, i.e. $r'_{\rm avg}=r'_{\rm clif}/2.2083=\left(9.42\pm0.09\right)\times10^{-5}$ as shown in Fig.~\ref{fig:pb_exp} (b), we estimate that the lower bound of the incoherent error contribution is $r_{\rm avg}^{\rm dec}/r'_{\rm avg} = 49.04\%$.

Besides the average gate fidelity and the error budget, experimentally-obtained process matrices undoubtedly provide more information on and insights into the implemented quantum operations, the underlying physical platform, and the involving control system. Here we introduce GST to obtain the process matrices of a single-qubit gate set ${\mathcal G}\equiv\left\{I, X_{\frac{\pi}{2}}, Y_{\frac{\pi}{2}}\right\}$. Compared to RB and QPT, the advantages of GST are of two folds. First, it provides process matrices of quantum operations in the same context, which are optimized simultaneously to fit the experimental outcomes. Next, the SPAM errors are separated from those in the noisy quantum gates by the introduction of the Gram matrix, i.e. a mathematical description of the SPAM. Finally, accuracy of $10^{-5}$ can be obtained with an experimentally feasible number of repetitions by using long and periodic gate sequences.

To facilitate the GST experiment and the following data analysis, we take advantage of the python package pyGSTi~\cite{nielsen2020QST}. The procedure begins with constructing a fiducial set ${\mathcal F}\equiv\{\emptyset, X_{\frac{\pi}{2}}, Y_{\frac{\pi}{2}}, X_{\frac{\pi}{2}}X_{\frac{\pi}{2}}X_{\frac{\pi}{2}}, Y_{\frac{\pi}{2}}Y_{\frac{\pi}{2}}Y_{\frac{\pi}{2}}, X_{\frac{\pi}{2}}X_{\frac{\pi}{2}}\}$, where $\emptyset$ denotes the null sequence. Acting on the initial state $\ket{0}$ and before the measurement on the $z$-basis, the operations in ${\mathcal F}$ span a symmetric and informationally-complete reference frame, in which arbitrary quantum operations can be unambiguously determined. To achieve Heisenberg-limited accuracy with respect to the circuit depth, a set of short circuits, i.e. the germs, should be elaborately chosen to amplify deviations in all parameters. The germ set contains 12 short circuits, i.e. $\{I, X_{\frac{\pi}{2}},Y_{\frac{\pi}{2}},X_{\frac{\pi}{2}}Y_{\frac{\pi}{2}},X_{\frac{\pi}{2}}X_{\frac{\pi}{2}}Y_{\frac{\pi}{2}},X_ {\frac{\pi}{2}}Y_{\frac{\pi}{2}}Y_{\frac{\pi}{2}},X_{\frac{\pi}{2}}Y_{\frac{\pi}{2}}I,X_{\frac{\pi}{2}}I\\Y_{\frac{\pi}{2}},X_{\frac{\pi}{2}}II,Y_{\frac{\pi}{2}}II,X_{\frac{\pi}{2}}Y_{\frac{\pi}{2}}Y_{\frac{\pi}{2}}I,X_{\frac{\pi}{2}}X_{\frac{\pi}{2}}Y_{\frac{\pi}{2}}X_{\frac{\pi}{2}}Y_{\frac{\pi}{2}}Y_{\frac{\pi}{2}}\}$, each of which is then repeated up to a length in the logarithmically spaced set $L\in\left\{1, 2, 4,\ldots, 4096\right\}$ to construct periodic gate sequences. The maximum depth is chosen such that, together with the repetition time for each circuit being $M=1024$, it leads to an estimated accuracy of $\sim\frac{1}{\sqrt{M}L}<10^{-5}$. The collected experimental data are analysed by the maximum likelihood estimation, leading to process matrices for the gate set which maximizes the log-likelihood function. The reconstructed process matrices are shown in Fig.~\ref{fig:gst_mv_mat} (a). With these matrices, we simulate the gate sequences used in the RB experiment shown in Fig.~\ref{fig:rb_exp} and obtain the simulated EPG $r^{\rm{sim}}_{\rm{avg}} = (8.97\pm0.03)\times10^{-5}$, as shown in Fig. \ref{fig:gst_mv_mat} (b). Note that these experiments, i.e. RB and GST, are executed in a relatively long period of time. The consistency of these two values shows the stability of our superconducting device, which is capable of implementing high-fidelity quantum gates over a relatively long period of time, while the deviation between them gives an intuitive quantification of the temporal fluctuation, which is on the level of $\sim 10^{-5}$.

Besides the estimated process matrices, the GST experiment also reveals the temporal correlation property, i.e. the non-Markovianity, of the physical platform in terms of model violations~\cite{nielsen2021Q}. The GST models a noisy quantum device with complete positive and trace preserving quantum channels, which are expressed by process matrices. Model violations emerge when there are deviations between the probabilities predicted by the optimal process matrices obtained from the GST and the observed frequencies in experiments, probably due to the existence of time-correlated errors, e.g. the slow drifting of system parameters. In pyGSTi, model violation is quantified by the loglikelihood score, which is operational defined as the distance between the optimized log-likelihood function and its mean, measured in the unit of its standard deviation, with the statistics given by the $\chi^2$-districution widely used in hypothesis tests. As to the GST experiment carried in our device, the individual loglikelihood scores (see SM~\cite{supplementary}) show statistically significant model violation mostly in long sequences with depth $\geq 256$. In other words, the non-Markovian effect can be measured on the accuracy level of $\sim10^{-4}$, which is comparable to the average EPG given by the RB experiment.

\section{\label{sec:level3}Discussion}

Although the non-Markovian errors defy accurate and reliable error analysis of the benchmarking results, direct monitoring of the control parameters still provides useful information about the error sources of the experimental platform. We consider two of the possible noise sources, i.e. classical noise from the electronics in the control system and the fluctuation of the transmon frequency, where the former corresponds to the AWG amplitude fluctuation in our experiment. These two fluctuations are monitored to be about $0.3\%$ and $0.1$ MHz (see SM~\cite{supplementary}), of which the contributions to the EPG are estimated to be $0.2\times10^{-5}$ and $0.1\times10^{-5}$, respectively. Specifically, we numerically simulate the evolution of the Schr\"odinger equation, with the pulse amplitude or the qubit frequency randomly sampled from normal distributions, whose mean and variance are determined by the target values and the monitored fluctuation strengths. As the estimated contributions to the EPG are much lower than the coherent part of the measured EPG, we conclude that the non-Markovianity is the main obstructive factor to further improve the fidelity.


\begin{acknowledgments}
This work is supported by the NSFC of China (Grants No. 11890704, 12004042, 11674376), the NSF of Beijing (Grants No. Z190012), National Key Research and Development Program of China (Grant No. 2016YFA0301800) and the Key-Area Research and Development Program of Guang-Dong Province (Grants No. 2018B030326001).
\end{acknowledgments}


%

\end{document}